\documentclass{JHEP3}
 \usepackage{epsfig}

 \title{
  On azimuthal spin correlations
  in Higgs plus jet events at LHC}

 \author{Kosuke Odagiri\\
  Institute of Physics, Academia Sinica, Nankang, Taipei, Taiwan 11529,
  The Republic of China}

 \abstract{
  We consider the recent proposal that the distribution of the difference
between azimuthal angles of the two accompanying jets in gluon-fusion
induced Higgs-plus-two-jet events at LHC reflects the CP of the Higgs
boson produced. We point out that the hierarchy between the Higgs boson
mass and the jet transverse energy makes this observable vulnerable to
logarithmically enhanced higher-order perturbative corrections.
  We present an evolution equation that describes the scale variation of
the azimuthal angular correlation for the two jets. The emission of extra
partons leads to a significant suppression of the correlation.
  Using the HERWIG Monte Carlo event generator, we carry out a
parton-shower analysis to confirm the findings.
  }

 \keywords{hig.jet.spe.hac}

 \preprint{hep-ph/0212215}

\begin{document}

 \section{Introduction}

  Gluon fusion via the top quark loop provides the dominant mechanism for
Higgs boson production at LHC.
  Recently, the analysis of the related $(2\to3)$ cross section was
carried out at the leading order in $\alpha_S(M_Z)$ in
ref.~\cite{delduca}.

  The proposal is to tag two extra jets, the motivation being the
elucidation of the differences/similarities between this process and the
weak-boson-fusion process, which also comes accompanied with two extra
jets.

  After imposing cuts on the jet momenta similar to the weak-boson-fusion
selection criteria, the soft-gluon contribution is reduced and they
obtain, as shown\footnote{We thank the authors of ref.~\cite{delduca} for
their kind permission to reproduce the figure.} in fig.~\ref{delducafig},
a striking correlation between the azimuthal angles $\phi$ of the two
jets, one of which is now in the forward direction while the other one is
in the backward direction. Provided that the produced Higgs boson is
CP-even, the distribution of the difference $\Delta\phi$ of the two
azimuthal angles is peaked at $\Delta\phi=0,\pi$ and falls to nearly zero
at $\Delta\phi=\pi/2$. If the Higgs boson is CP-odd, although the result
is not explicitly shown, the distribution is peaked at $\Delta\phi=\pi/2$
and falls to nearly zero at $\Delta\phi=0,\pi$.
 \FIGURE{
 \epsfig{file=phi_cuts_comp.eps, width=10cm}
 \caption{The distribution of the azimuthal angle between the two highest
$p_T$ jets, taken from ref.~\cite{delduca}. Results shown are for the
top-quark induced gluon-fusion process, with $m_t=175$ GeV and in the
limit $m_t\to\infty$, and for the weak-boson-fusion process.
 }
 \label{delducafig}
 }

  In this work, we would like to point out one potential pitfall which
seems to have been neglected in their study, namely that there are two
scales in this problem. The higher scale is related to the Higgs boson
production. For a leading order analysis we can set it to be the Higgs
boson mass for convenience. The lower scale is related to the emission
leading to the tagged jets and this is characterized by the jet transverse
momenta.
  Because of the presence of two scales, the predictions of calculations
at a finite perturbative order becomes sensitive to higher order
corrections.

  We investigate this problem by first establishing an evolution equation
that describes the scale variation of the azimuthal angular correlation
coefficient. The large size of the relevant anomalous dimension implies
that there is significant, up to one order of magnitude, reduction in the
size of the correlation coefficient.

  Although the problem is formally due to the logarithm of the ratio of
the two scales which enhance the higher order contributions, we find that
the ratio needs not be so large for the effect of extra emission to become
important.

  An alternative approach to investigating this problem is by a
parton-shower level Monte Carlo simulation. We formulate this problem as a
$(2\to1)$ cross section convoluted with the leading logarithmic parton
shower using HERWIG \cite{herwig}, which includes the azimuthal spin
correlations by default by using the algorithm of Collins and Knowles
\cite{collins, knowles, peter}.

  The results of the two approaches are in good agreement.

  We note that the weak-boson fusion mode is unaffected by extra emission,
as the jet $p_T$ scale is the only QCD scale present in this case.

  We present the evolution equation analysis in sect.~\ref{evolution}
  and its comparison with the parton-shower analysis in
sect.~\ref{partonshower}. We present the conclusions in
sect.~\ref{conclusions}.

 \section{Evolution equation analysis}\label{evolution}

  Let us consider the evolution of a jet that is due to an initial state
parton. In a Monte Carlo simulation, this is described by the backward
evolution.

  At each stage of evolution, the probed parton has virtuality $t$,
momentum fraction $x$, jet transverse momentum $p_T$ according to some
definition, and spin density $\rho$. We wish to measure the mean value of
the component of $\rho=\rho_\parallel$ that is aligned with the reference
direction given by $p_T$. The correlation arises in the first place
because of the correlation, at the hard process level, between the planes
of polarization of the gluons involved in the Higgs boson production.
Hence the decorrelation between the plane of polarization of each gluon
and the direction of the related tagged jet is equivalent to the
suppression of the azimuthal angular correlation between the two tagged
jets.

  In the following, $p_T$ and $\rho$ are both two-component vectors. The
spin density matrix $\rho_{ij}$ is given in terms of the vector $\rho_k$
by $\rho_{ij}=(1+\rho_k\sigma^k_{ij})/2$. The third component of the
vector $\rho$ vanishes so long as the nucleon is unpolarized
\cite{collins, knowles}.
  In terms of the distribution functions $f(x,t,p_T,\rho)$, we may write
the scale variation as follows:
 \begin{eqnarray}
  \frac{\partial<\!\rho_\parallel\!>\!(x,t)}{\partial t}&=&
  \frac\partial{\partial t}
  \frac{\int f(x,t,p_T,\rho) \rho_\parallel dp_Td\rho}
  {\int f(x,t,p_T,\rho) dp_Td\rho} \nonumber\\&=&
  \frac1{f(x,t)}\int\frac{\partial f(x,t,p_T,\rho)}{\partial t}
  \{\rho_\parallel-<\!\rho_\parallel\!>\!(x,t)\}
  dp_Td\rho. \label{master_eqn}
 \end{eqnarray}

  The change in the distribution functions when the scale is raised from
$t$ to $t+\delta t$ is given by:
 \begin{eqnarray}
 \delta f(x,t,p_T,\rho) &=& \delta f_\mathrm{in}-\delta f_\mathrm{out},
 \label{delta_f}\\
 \delta f_\mathrm{out} &=&
 \frac{\delta t}t\int dz\frac{\alpha_S}{2\pi}\frac{d\phi}{2\pi}
 CF(z,\phi,\rho) f(x,t,p_T,\rho), \label{f_out}\\
 \delta f_\mathrm{in}  &=&
 \frac{\delta t}t\int \frac{dz}z\frac{\alpha_S}{2\pi}\frac{d\phi}{2\pi}
 d\rho' CF(z,\phi,\rho') \delta(\rho-\rho(\rho',z,\phi))
 \nonumber\\&&
 \times\ f(x/z,t,p_T'(p_T,t,z,\phi),\rho').\label{f_in}
 \end{eqnarray}
  In eqn.~(\ref{f_in}), $\phi$ is the emission angle. $z$ is the remaining
energy fraction during the emission, such that the emitted energy fraction
is $1-z$.
  $\rho$ is determined by the following relation according to the notation 
of ref.~\cite{knowles}:
 \begin{equation}
  \rho=(f_2(z)n(\phi)+f_3(z)\rho')/F(z,\phi,\rho'), \label{newrho}
 \end{equation}
  where $f_i$ are various components of the splitting function shown in
tab.~\ref{splitting}, and $n(\phi)=(\cos2\phi,\sin2\phi)$. In all of the
above, $C$ and $F$ are the colour factor and the polarization dependent
splitting function respectively, with $F(z,\phi,\rho')=\widehat
P(z)/C+f_4(z)n(\phi)\!\cdot\!\rho'$.
 \TABLE{
  \begin{tabular}{ccccc}
  \hline
  Process & $\widehat P/C$ & $f_2$ & $f_3$ & $f_4$ \\
  \hline
  $g\to gg$ & $2\left[(1-z)/z+z/(1-z)+z(1-z)\right]$ &
  $2(1-z)/z$ & $2z/(1-z)$ & $2z(1-z)$       \\
  $g\to q\bar q$ & $z^2+(1-z)^2$ & $0$ & $0$ & $-2z(1-z)$ \\
  $q\to qg$ & $(1+z^2)/(1-z)$ & $0$ & $2z/(1-z)$ & $0$ \\
  $q\to gq$ & $\left[1+(1-z)^2\right]/z$ & $2(1-z)/z$ & $0$ & $0$ \\
  \hline
  \end{tabular}
  \label{splitting}
  \caption{The coefficients $f_i$, taken from ref.~\cite{knowles},
modified to adopt the convention of ref.~\cite{esw} for the 
normalization.}
  }

  Combining the above with eqn.~(\ref{master_eqn}), the contribution from
the term $\delta f_\mathrm{out}$ vanishes and we obtain:
 \begin{eqnarray}
  \frac{\partial<\!\rho_\parallel\!>\!(x,t)}{\partial t}&=&
  \frac1{f(x,t)}
  \int\frac{\partial f_\mathrm{in}}{\partial t}
  \left(\rho_\parallel-<\!\rho_\parallel\!>\!(x,t)\right)
  dp_Td\rho \\ &=&
  \frac{1/t}{f(x,t)}\int \frac{dz}z\frac{\alpha_S}{2\pi}\frac{d\phi}{2\pi}
  d\rho' dp_Td\rho CF(z,\phi,\rho')
  \delta(\rho-\frac{f_2(z)n(\phi)+f_3(z)\rho'}{F(z,\phi,\rho')})
  \nonumber\\ && \times
  f(x/z,t,p_T'(p_T,t,z,\phi),\rho')
  \left(\rho_\parallel-<\!\rho_\parallel\!>\!(x,t)\right)
  \label{eqn1}\\ &=& -\frac1t\int
  \frac{dz}z\frac{\alpha_S}{2\pi}\frac{d\phi}{2\pi}
  d\rho' dp_T' C \frac{f(x/z,t,p_T',\rho')}{f(x,t)} \nonumber\\ &&
  \left[<\!\rho\!>\!\!(x,t)
  F(z,\phi,\rho')-(f_2(z)n(\phi)+f_3(z)\rho')\
  \right]\!\cdot\! (\cos2\phi_0,\sin2\phi_0). \label{eqn2}
 \end{eqnarray}
  In the last line, $\phi_0$ is the azimuthal angle of $p_T$. Between
eqns.~(\ref{eqn1}) and (\ref{eqn2}), we have assumed that $p_T$ is defined
in an additive manner, such that $dp_T=dp_T'$. Excepting this assumption,
our discussion so far has been general.

  Now let us consider the case in which the direction of $p_T$, or more
generally the reference direction, is to a good approximation determined
at one scale given, for instance, by the jet transverse momentum
${p_T}_j$, and we are interested in the soft emission
  that depolarizes the gluon
  between this scale and the hard process scale.

  In this case, $\phi_0$ can be taken as constant. Integrating over
$\phi$, by symmetry, the term proportional to $n(\phi)$ in
eqn.~(\ref{eqn2}) vanishes, as does the $\phi$ dependent term in
$F(z,\phi,\rho')$. We then have:
 \begin{eqnarray}
  \frac{\partial<\!\rho_\parallel\!>\!(x,t)}{\partial\ln t}&=&
  -\int\frac{dz}z\frac{\alpha_S}{2\pi} 
  d\rho' dp_T' \frac{f(x/z,t,p_T',\rho')}{f(x,t)} \nonumber\\ &&
  \left[\widehat P(z)<\!\rho_\parallel\!>\!(x,t)
  -Cf_3(z)\rho_\parallel' \right] \label{eqn3}\\&=&
  -\int\frac{dz}z\frac{\alpha_S}{2\pi}
  \frac{f(x/z,t)}{f(x,t)} \nonumber\\ &&
  \left[\widehat P(z)<\!\rho_\parallel\!>\!(x,t)
  -Cf_3(z)<\!\rho_\parallel\!>\!(x/z,t) \right]. \label{eqn4}
 \end{eqnarray}

  The pole at $z\to1$ of $\widehat P(z)$ cancels with the pole of
$f_3(z)$. The expression of eqn.~(\ref{eqn4}) is the difference between
the change in the structure function $f(x,t)$ given by the first term and
the change in the spin density $<\!\rho_\parallel\!>f(x,t)$ given by the
second term. The two terms can be treated separately by introducing the
plus prescription to account for the $\delta f_\mathrm{out}$ term given by
eqn.~(\ref{f_out}). We obtain:
 \begin{eqnarray}
  \frac{\partial<\!\rho_\parallel\!>\!(x,t)f(x,t)}{\partial\ln t}&=&
  \int\frac{dz}z\frac{\alpha_S}{2\pi}
  f(x/z,t)<\!\rho_\parallel\!>\!(x/z,t)\ Cf_3(z)_+, \label{eqn5}\\
  \frac{\partial f(x,t)}{\partial\ln t}&=&
  \int\frac{dz}z\frac{\alpha_S}{2\pi}
  f(x/z,t) P(z). \label{eqn6}
 \end{eqnarray}

  Taking the $x^j$ moments of the above to obtain Mellin transforms, we
convert the expressions into:
 \begin{eqnarray}
  \frac{\partial \widetilde{\left[<\!\rho_\parallel\!>f\right]}
  (j,t)}{\partial\ln t}&=&
  \widetilde{\left[<\!\rho_\parallel\!>f\right]}(j,t)
  \int dzz^{j-1}\frac{\alpha_S}{2\pi} Cf_3(z)_+, \label{eqn7}\\
  \frac{\partial \widetilde f(j,t)}{\partial\ln t}&=&
  \widetilde f(j,t)
  \int dzz^{j-1}\frac{\alpha_S}{2\pi} P(z), \label{eqn8}
 \end{eqnarray}
  such that the anomalous dimensions are:
 \begin{eqnarray}
  \gamma_{\!<\!\rho_\parallel\!>\!f}
   &=& \int dzz^{j-1}\frac{\alpha_S}{2\pi} Cf_3(z)_+, \label{eqn9}\\
  \gamma_{f}
   &=& \int dzz^{j-1}\frac{\alpha_S}{2\pi} P(z), \label{eqn10}
 \end{eqnarray}
  respectively for the spin density and the parton distribution functions.

  As stated above, eqn.~(\ref{eqn4}) specifies the evolution of
polarization in regions where extra emission does not alter the reference
direction. For our purpose, the most interesting case is the
soft/collinear gluon emission from the gluon line in between the hard
process and the jet $p_T$ scale.

  For small enough $x$, such that $j\sim1$, the behaviour of the anomalous
dimensions is controlled by the $z\to0$ region.
  If we choose the $g\to gg$ splitting, the fact that $P(z)$ has a pole at
$z\to0$ where as $f_3(z)_+$ does not, indicates that a large disparity
arises in the two quantities.
  $<\!\rho_\parallel\!>$ is the ratio of $<\!\rho_\parallel\!>\!f$ and
$f$, such that its running is governed by the difference of the above two
anomalous dimensions. The relevant splitting function is:
  \begin{equation}
  \left[Cf_3(z)-\widehat P(z)\right]_{g\to gg}=-2C_A
  \frac{(1-z)(1+z^2)}z.
  \end{equation}
  Exactly as in the case of the hadron multiplicity calculation
\cite{mueller,esw}, the expression of eqn.~(\ref{eqn10}), when $j=1$, is
divergent as written, but a more careful analysis, taking into account the
suppression of soft gluon emission due to coherence, leads to the result:
 \begin{eqnarray}
  \Delta\gamma(j,\alpha_S) &=&
  -\sqrt{\frac{C_A\alpha_S}{2\pi}}+\frac{j-1}4+\ldots. \\
  \frac{\widetilde{\left[<\!\rho_\parallel\!>f\right]}(1,t)}
       {{\widetilde f}(1,t)}
  &\propto& \exp\left[-\frac1b\sqrt\frac6{\pi\alpha_S(t)}\right]
  \sim \exp
  \left[-\sqrt{\frac6{\pi b}\ln\left(\frac{t}{\Lambda^2}\right)}\right].
  \label{eqn11}
 \end{eqnarray}
  Hence the evaluation of $<\!\rho_\parallel\!>$ at the hard process scale
$\sim\sqrt{\hat s}$, defining it to be the ratio of
$<\!\rho_\parallel\!>\!f$ and $f$ individually integrated over the
allowed, typically small, values of $x$, assuming perfect polarization at
the lower scale $\sim{{p_T}_j}$, gives:
 \begin{equation}
  <\!\rho_\parallel\!>({p_T^2}_j) \approx \exp \left[
  \sqrt{\frac6{\pi b}\ln\frac{{p^2_T}_j}{\Lambda^2}} -
  \sqrt{\frac6{\pi b}\ln\frac{\hat s}{\Lambda^2}} \right]
  \label{eqn12}
 \end{equation}

  Let us take the jet energy scale at around 20 GeV and the hard process
scale at 100 GeV. $b=(11C_A-2n_f)/12\pi$ at this order, and we can adopt
for simplicity $\Lambda\sim200$ MeV. We then obtain $<\!\rho_\parallel\!>
= 0.42$. We are interested in the correlation between two jets, which is
now given by the square of $<\!\rho_\parallel\!>$, and we obtain $0.18$.
  Thus there is almost one order of magnitude dilution.

  There would be further dilution due to the modification of the $p_T$
direction due to extra emission, and the imperfect polarization at the jet
energy scale must also be taken into account. The parton-shower analysis,
which we present in the following, incorporates all of these effects.

 \section{Parton-shower simulation}\label{partonshower}

  Our parton-shower simulation is based on the Monte Carlo event generator
HERWIG 6.5 \cite{herwig}. For the $(2\to1)$ hard process,
 \begin{equation}
  gg\to \Phi, \label{ggtoPhi}
 \end{equation}
  where $\Phi$ is either a CP-even or a CP-odd Higgs boson with
$M_\Phi=120$ GeV \cite{delduca}, we utilize the code for the MSSM $H^0$
and $A^0$ production. We set $\tan\beta=1$, $\beta-\alpha=\pi/2$ and all 
supersymmetric masses equal to 10 TeV, such that the Higgs production is 
mostly due to the top quark loop only.
  The parton shower, including the azimuthal angular correlation
\cite{collins, knowles}, follows the HERWIG default, although we found two
bugs which had to be corrected before the effect of the angular
correlation could be seen. One bug affects the azimuthal angle of the
branching $q\to gq$, whereas the other, more serious, bug affects the
Lorentz boost of the jets to the laboratory frame. The resultant code will
be part of the next HERWIG sub-version.

  We have performed the analysis also incorporating the heavy quark
initiated hard process $Q\bar Q\to\Phi$, but there was no significant
alteration to the results presented herein.

  For jet reconstruction, we make use of the program GETJET \cite{getjet}.
This uses a simplified version of the UA1 jet algorithm with jet radius
$\Delta R$ which we set equal to 0.4.

  For the Standard Model parameters, we adopted the default values in
HERWIG.
  For the structure function, we have used the default set in HERWIG,
namely the mean of the central gluon and higher gluon leading order
structure functions of MRST98 \cite{mrst98lo}.
  We follow the numbers in ref.~\cite{delduca} for the Higgs boson mass of
120 GeV and the jet selection criteria. We do not decay the Higgs boson
and do not impose any cuts on its kinematics. We take the LHC 
centre-of-mass energy $\sqrt{s}=14$ TeV.

  The jet selection criteria, which we define to be applicable to the two
jets with the highest transverse momenta ${p_T}_j$, are as follows:
 \begin{eqnarray}
 && {p_T}_j > 20\ \mathrm{GeV},\quad
 |\eta_j| < 5,\quad
 \Delta R_{jj} > 0.6 \nonumber\\
 && |\Delta\eta_{jj}| > 4.2,\quad
 \eta_{j1}\cdot\eta_{j2}<0,\quad
 m_{jj}>600\ \mathrm{GeV}.
 \label{jetcuts}
 \end{eqnarray}
  Thus we are picking out events with one forward jet and one backward
jet, with large combined invariant mass.

 \FIGURE{
 \epsfig{file=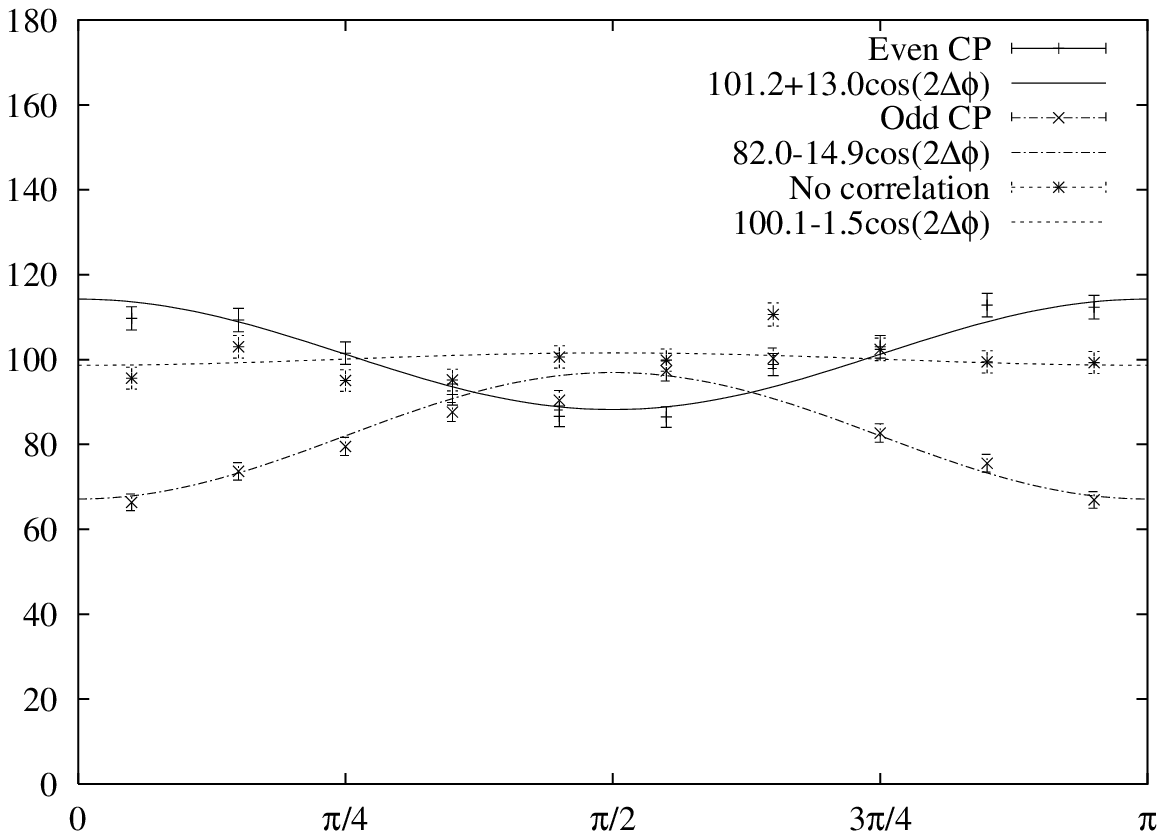, width=10cm}
 \put(-293,203){$\frac{d\sigma}{d\Delta\phi}\ /\mathrm{fb}$}
 \put(-143,-10){$\Delta\phi$}
 \caption{The distribution of the azimuthal angle between the two highest
$p_T$ jets, after imposing the cuts of eqn.~(3.2). 
Simulation at the parton-shower level. The error-bars are for the
statistics. }
 \label{partonlevel}
 }

  The result of our parton-shower simulation on the azimuthal angular
difference $\Delta\phi$, following the procedures outlined above, is shown
in fig.~\ref{partonlevel}. The error-bars are for the statistics. We
generated one million events in each case to compensate for the low
acceptance.
  Since the total cross section is around 20 pb, this corresponds to about
50 fb$^{-1}$ integrated luminosity.

  The difference in the normalization between the CP-even case and the
CP-odd case is due to the different loop amplitudes depending on whether
the produced Higgs boson is CP-even or CP-odd. The total cross section is
similar to that shown in fig.~\ref{delducafig}. The points marked `No
correlation' corresponds to the same hard process as the CP-even case but
with the azimuthal correlation turned off by means of the HERWIG option
{\tt AZSPIN=.FALSE.}

  The curves in fig.~\ref{partonlevel} are fitted by the Fourier series
analysis using the expansion $a_0+a_1\cos(2\Delta\phi)$. Our fit gives
$101.2+13.0\cos(2\Delta\phi)$ (fb) for the CP-even case and
$82.0-14.9\cos(2\Delta\phi)$ (fb) for the CP-odd case. With the azimuthal
correlation switched off, we obtain $100.1-1.45\cos(2\Delta\phi)$ (fb) so
that it is possible that the case without azimuthal correlation is not
completely flat. We calculated the statistical error on the Fourier
coefficients to be about 1 fb for all coefficients. The coefficient in
this case is found to be $1.24\sigma$ away from being flat.

  The correlation is evidently diluted. As the distribution is almost
proportional to $1+\cos(2\Delta\phi)$ for the CP-even case in
fig.~\ref{delducafig}, we may say that there is dilution by nearly one
order of magnitude, in agreement with the analysis in
sect.~\ref{evolution}.

  Next, let us look at the jet $p_T$ scale dependence of the correlation
coefficient. In order to account for the possibly intrinsically non-flat
distribution, we calculate the mean of the CP-even and CP-odd correlation
coefficients given by:
 \begin{equation}
  \left<\left|\frac{a_1}{a_0}\right|\right>
  = \frac12\left[\left(\frac{a_1}{a_0}\right)_\mathrm{CP-even}-
    \left(\frac{a_1}{a_0}\right)_\mathrm{CP-odd}\right].
 \end{equation}
  For the evolution equation analysis, the correlation coefficient is
calculated as the square of $<\!\rho_\parallel\!>$ given in
eqn.~(\ref{eqn12}).

 \FIGURE{
 \epsfig{file=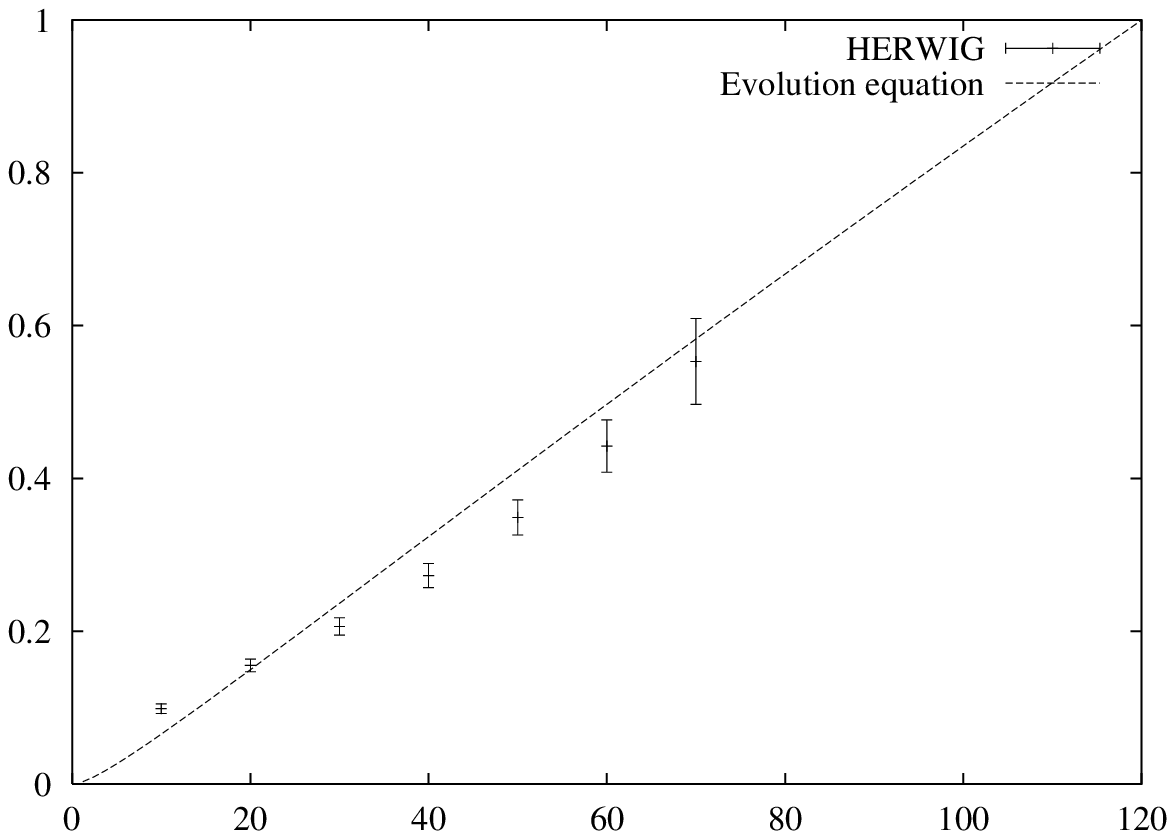, width=10cm}
 \put(-293,203){$\left<\left|\frac{a_1}{a_0}\right|\right>$}
 \put(-180,-10){Jet $p_T$ scale /GeV}
 \caption{Jet $p_T$ scale dependence of the correlation coefficient
$|a_1/a_0|$, averaged over the CP-even and CP-odd cases. For the HERWIG
numbers, the jet $p_T$ scale is taken to be the minimum ${p_T}_j$, whereas
for the evolution equation analysis, the jet $p_T$ scale is as represented
by ${p_T}_j$ in eqn.~(2.20). The error-bars on the HERWIG numbers are for
the statistics.
 }
 \label{evolutionplot}
 }
  The result is shown in fig.~\ref{evolutionplot}, taking the jet $p_T$
scale to be the minimum ${p_T}_j$ for the HERWIG simulation and ${p_T}_j$
in the evolution equation analysis as seen in eqn.~(\ref{eqn12}). We
rather arbitrarily chose $\Lambda=180$ MeV in this case, equal to the
value of the HERWIG variable {\tt QCDLAM}. We set the higher scale
$\sim\sqrt{\hat s}=120$ GeV to correspond to the Higgs boson mass.

  We see that the HERWIG results are in good agreement with the evolution
equation analysis in sect.~\ref{evolution}. The running of the correlation
coefficient at low jet energy scales is slower in HERWIG. This is natural
when we consider the fact that the actual ${p_T}_j$ is always higher than
the imposed ${p_T}_j$ cut, especially so when the cut is low. We verified
this by taking the ${p_T}_j>10$ GeV point and further imposing the
constraint that neither of the tagged jets have ${p_T}_j$ greater than 20
GeV. We obtain $0.000\pm0.024$ for the coefficient in this case.

  As mentioned in sect.~\ref{evolution}, there is further dilution in the
HERWIG analysis compared to the evolution equation analysis due to the
imperfect polarization at the jet $p_T$ scale and the modification of the
$p_T$ direction due to extra emission. The latter is expected to be
important when the jet $p_T$ scale is low. The result in
fig.~\ref{evolutionplot} indicates that although these effects must be
present, they do not significantly alter the behaviour derived by the
evolution equation analysis.

 \FIGURE{
 \epsfig{file=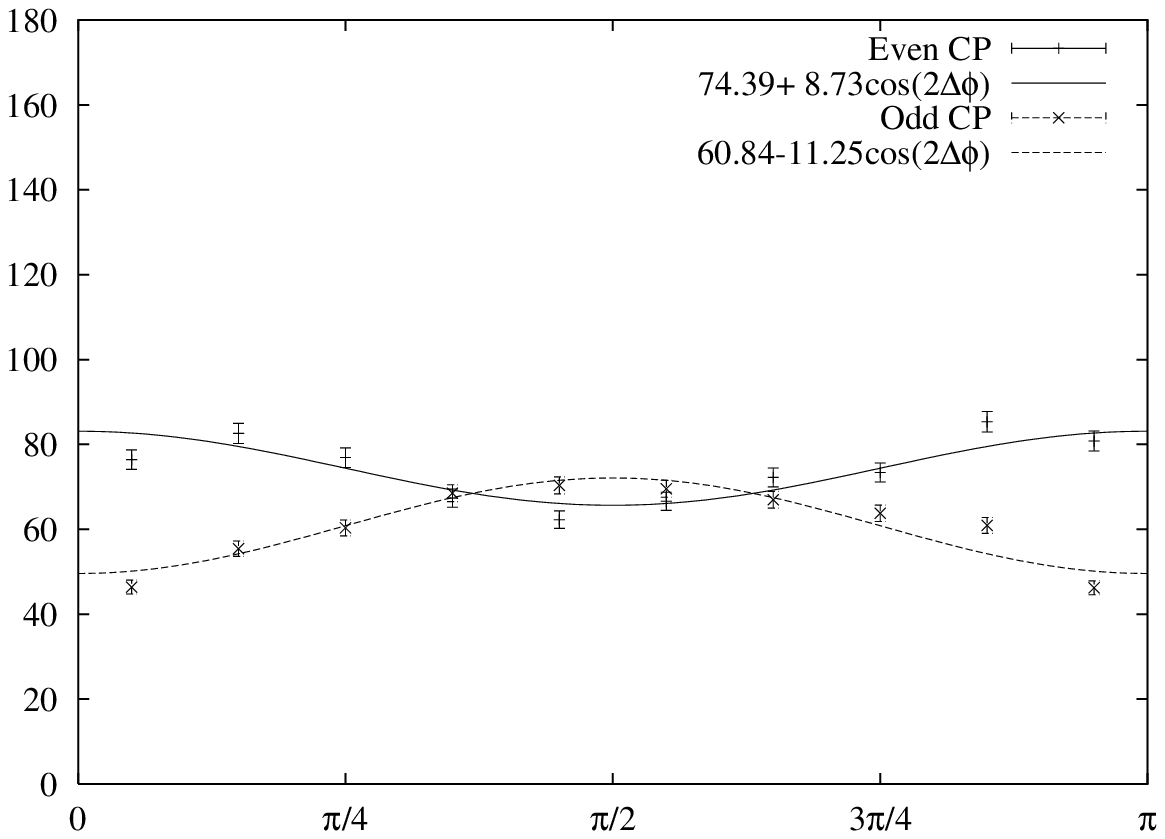, width=10cm}
 \put(-293,203){$\frac{d\sigma}{d\Delta\phi}\ /\mathrm{fb}$}
 \put(-143,-10){$\Delta\phi$}
 \caption{The distribution of the azimuthal angle between the two highest
$p_T$ jets, after imposing the cuts of eqn.~(3.2). 
Simulation at the hadron level, without soft underlying events.
 }
 \label{hadronlevel}
 }
  Our analysis so far has been carried out at the parton-shower level.
  Let us now turn our attention to the effect of hadronization.
  In fig.~\ref{hadronlevel}, we show the numbers for the hadron level
simulation corresponding to the cuts of eqn.~(\ref{jetcuts}), i.e., with
${p_T}_j>20$ GeV. We have turned off soft underlying events. We have
utilized the simplified calorimeter simulation in GETJET. For the
electromagnetic calorimeter, we take the resolution to be
$\sigma_E/E=10\%/\sqrt{E/\mathrm{GeV}}$, and for the hadronic calorimeter,
we take the resolution to be $\sigma_E/E=50\%/\sqrt{E/\mathrm{GeV}}$.
  We observe that except for the reduced acceptance rate, the result is
similar to that shown in fig.~\ref{partonlevel}. The resulting coefficient
$<|a_1/a_0|>$ is consistent with the parton level simulation.

 \section{Conclusions}\label{conclusions}

  By means of an evolution equation analysis and the HERWIG parton
shower, we have performed an all-order study of the jet azimuthal angular
distribution of events containing a Higgs boson and at least two jets at
LHC.

  The results of the two analyses are consistent with each other and
predict that due to the emission of extra partons, the azimuthal angular
correlation that arises in the fixed-order analysis is diluted by almost
one order of magnitude.

  Although we have specialized to the case of Higgs boson production, our
results are applicable to jets accompanying any hard process initiated by
two gluons.

  We have verified that hadronization does not significantly modify our
conclusions.

 \subsection*{Acknowledgements}

  This work was at one stage intended as a collaborative study with
Yoshiaki~Yasui and Eri~Asakawa. I am very much grateful to them for
discussions and their interest in this work.

  I thank Bryan Webber for penetrative comments and guidance. I have also
benefited from discussions with, and encouragement from, many others,
including Ian~Knowles, Mike~Seymour, Tim~Steltzer, and the members of the
KEK theory group and the HEP group of the Institute of Physics,
Academia~Sinica.

  I thank the authors of ref.~\cite{delduca} for reading and commenting on
the manuscript, and for their permission to reproduce
fig.~\ref{delducafig}.

 \end{document}